# Non-dipole effects in two-photon sweeping of the *K*-shell of an atomic ion

A.N. Hopersky, A.M. Nadolinsky, S.A. Novikov, *and* R.V. Koneev

Rostov State Transport University, 344038, Rostov-on-Don, Russia
E-mail: qedhop@mail.ru, amnrnd@mail.ru, sanovikov@gmail.com, koneev@gmail.com

**Abstract.** In the work [1], within the framework of a *dipole* approximation for the radiation transition operator, of the second-order nonrelativistic quantum perturbation theory and the Hartree-Fock single-configuration approximation, the first theoretical study of the generalized cross-section of the *direct* two-photon sweeping out of the *K*-shell of a light neon atom was carried out. **In this Letter** we supplement the Preprint [2] with the results of taking into account *non-dipole* effects when constructing the amplitude of the probability of the radiation transition between *continuum*-spectrum states. As the main result, it was found that the *non-dipole* effects reduce the generalized cross-sections of two-photon sweeping out of the *K*-shell of an atomic ion ($Fe^{16+}$) the calculated in the *dipole* approximation by several orders of magnitude (*giant* non-dipole effect).

Without repeating the content of the Preprint [2], we (keeping the numbering of the claims) will reproduce and supplement only the results of Sec. 2.4 and Sec. 3 for the partial generalized cross-sections (probabilities of photons disappearance without photoelectrons registration):

$$\sigma_g^{(pp)} \cong \frac{2\pi}{\gamma_{1s}} \cdot \mu \cdot [\omega + I(1s^{-2}) - 2I_{1s}]^2 \cdot \left(\frac{4}{3} - \frac{1}{4\pi}\right) \cdot L^2, \tag{41}$$

$$\mu = \frac{8\pi^3}{9V} \cdot \alpha \cdot r_0^2 (a_0 c\hbar)^2 = 0.278 \cdot 10^{-52} \text{ [cm}^4\text{·s]}, \tag{42}$$

$$L = N_{sp} \langle 1s_0 \| \hat{r} \| \varepsilon p_+ \rangle \langle 1s_+ | \hat{r} | \varepsilon p_{++} \rangle, \tag{43}$$

$$\sigma_g^{(ss)} = \mu K^2 \cdot J, \tag{44}$$

$$K = N_{sp} \cdot \frac{\omega \sqrt{2\varepsilon}}{\gamma_{1s}^2} \langle 1s_0 \| \hat{r} \| \varepsilon p_+ \rangle, \tag{45}$$

$$J = \frac{a}{3} \cdot [8f^2(a/2) + (f(0) + f(a))^2], \tag{46}$$

$$\sigma_g^{(sd)} = \frac{6}{5}\left(1 - \frac{1}{4\pi}\right) \cdot \sigma_g^{(ss)}, \tag{47}$$

where $\alpha$ is the fine-structure constant, $r_0$ is the classical radius of the electron, $a_0$ is the Bohr radius and $a = 2\omega - I(1s^{-2})$.

Let's establish an analytical structure of the function *K* from (45) *outside* the *dipole* approximation. The amplitude of the transition probability (28) from [2] is obtained in a *dipole* approximation for the radiation transition $\hat{R}$-operator. For the matrix elements $\langle xp_+ | \hat{r} | z(s,d)_{++} \rangle$ this approximation, strictly speaking, is *incorrect*. In fact, in this case, there **is no concept** of a "localization region" of the wave functions of transition states, which is less than the wavelength of the absorbed photon. Going beyond the dipole approximation is carried out through the decomposition of the operator $\hat{A}_n$ in (7) from [2] an infinite functional series over *multipoles* [3-5]. As a result of the transformation of this decomposition [in the integral representation of the multipole Hamiltonian interaction [5] (*Chapter* 5, § 5.2) via communication of the electric field operator with the electromagnetic field operator $\hat{E} = -(1/c)\partial \hat{A}/\partial t$ at *t* = 0] we assumed the approximation of the works [6, 7] for a single-electron operator of the radiation transition between *continuum*-spectrum states:

$$\hat{r}, r \in [0; \infty) \to \hat{D} = \begin{cases} \hat{r}, r \in [0; q), \\ q, r \in [q; \infty), \end{cases} \quad (48)$$

where the magnitude $q = (3/8)\lambda_\omega$ depends only on the energy of the absorbed photon and turns out to be the «*stopping point*» of the *infinite* growth of the operator $\hat{r}$. Then, after substituting (48) to (28) from [2], the function $K$ is modified by substituting:

$$K \to K\Psi(\omega), \quad (49)$$

$$\Psi(\omega) = 1 - \exp\{-(\gamma_{1s}/\sqrt{2\varepsilon})q\}, \quad (50)$$

The calculation results are shown in Figs. 2,3. For the parameters of generalized cross-sections, the following values are taken: $\Gamma_{1s} = 1.046$ eV, $I_{1s} = 7699.23$ eV, $I(1s^{-2}) = 15811.77$ eV and $\hbar\omega \in (6; 100)$ keV.

According to Fig. 2, *outside* the dipole approximation the following order of magnitude of the ratio of the generalized cross-sections of the two-photon sweeping of the $K$-shell is obtained:

$$(\sigma_g^{(ss)} + \sigma_g^{(sd)})/\sigma_g^{(pp)} \cong 10^6. \quad (51)$$

The result (51) in the formalism of Feynman diagrams can be given the following physical interpretation. According to Fig. 1b, the first photon at a time $t_1$ ionizes the $K$-shell into a *virtual* $xp_+$–state of the continuous spectrum. The second photon is absorbed at a moment of time $t_2 > t_1$ by the "cloud" of the $xp_+$–continuous spectrum. The $z(s,d)_{++}$–continuum-spectrum state produced in this case by Coulomb repulsion ejects the remaining $1s$-electron of the $K$-shell into the $ys_{++}$–continuum state of the same symmetry (see overlapping integral $\langle 1s_+ | ys_{++} \rangle$). Such a process is much more likely than the process of free passage of the second photon through the "cloud" and its subsequent absorption by the remaining $1s$–electron of the $K$-shell (Fig. 1a).

According to Fig. 3, for $\omega \geq I(1s^{-2})/2$ *outside* the dipole approximation the following order of magnitude of the ratio of the generalized cross-sections of two-photon sweeping and two-photon single ionization of the $K$-shell obtained:

$$(\sigma_g^{(ss)} + \sigma_g^{(sd)} + \sigma_g^{(pp)})/(\sigma_g^{(s)} + \sigma_g^{(d)}) \cong 10^3. \quad (52)$$

The result (52) can be given the following physical interpretation. The probability of the $z(s,d)_{++}$ –continuum-spectrum state of ejecting the remaining $1s$–electron of the $K$-shell into the $ys_{++}$– state of the continuous spectrum (Fig. 1b) is higher than the probability of its ($1s$–electron) not being "noticed" (see background section in Fig. 3).

The result (52) qualitatively reproduces that of the work [1] for the atom Ne. At the same time, we note the following. In [8], the generalized cross-section of the two-photon double ionization of the $K$-shell of the atom Ne $\sigma_g \sim 10^{-53}$ (cm$^4$·s) at $\omega \approx 1$ keV was estimated, which is much less than the value $\sigma_g \sim 10^{-49}$ (cm$^4$·s) of the work [1]. The generalized cross-section of the work [1] is obtained in the *dipole* approximation for the $\hat{R}$-operator. Taking into account *non-dipole* effects [see (49)] leads to agreement with the result of the work [8]. At the same time, in [8] there is no information (for a *detailed* comparison with the mathematical formalism of the work [1]) on the method of accounting for the matrix elements of the $\hat{R}$-operator with the transition states of the *continuous* spectrum. It can be assumed that the approximation (48) of the works [6, 7] was *first* realized for the theoretical description of the effect of two-photon *sweeping* of the $K$-shell of an atomic ion.

Thus, a *giant* (see Figs. 2b, c and Fig. 3) non-dipole effect was found in the calculation of generalized cross-sections from the $ss$ ($^1S_0$)– and $sd$ ($^1D_2$)–channels of double ionization of the $K$-shell (see Fig. 1b).

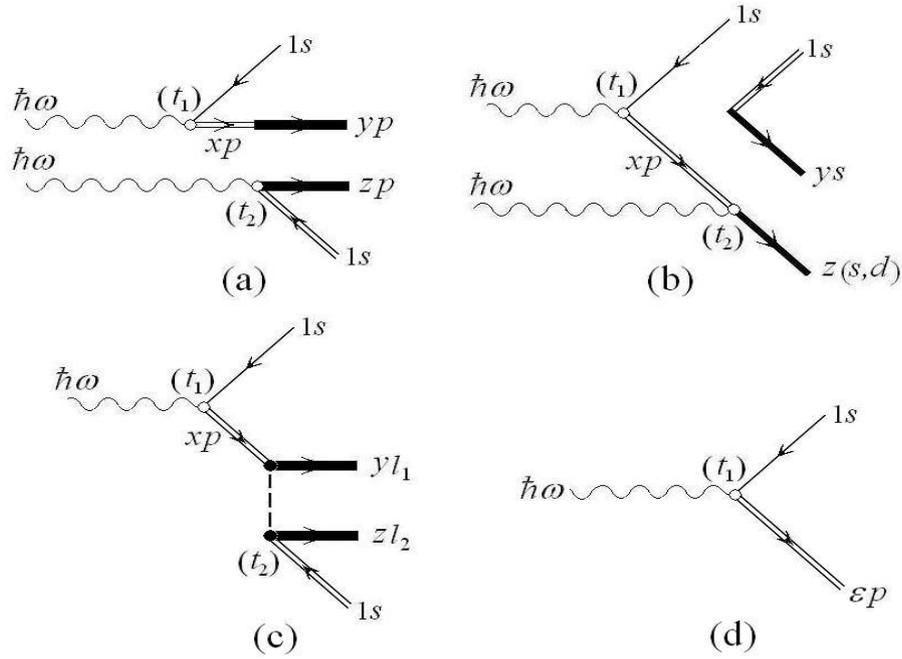

**Figure 1.** Probability amplitude of two-photon and single-photon ionization of the *K*-shell of a neon-like atomic ion ($Fe^{16+}$) in the representation of (non-relativistic) Feynman diagrams: (a) on channel (1) from [2] for $l_1 l_2 = pp$; (b) by channels (1) from [2] for $l_1 l_2 = ss, sd$; (c) over the channels $\hbar\omega + [0] \to 1sxp \to 1s^0 yl_1 zl_2$ for $l_1 l_2 = sp, pd, df, \ldots$; (d) on the channel $\hbar\omega + [0] \to 1s\varepsilon p$, $\varepsilon = \hbar\omega - I_{1s}$. An arrow to the left is a vacancy, an arrow to the right is an electron. Double (wide black) line – the states are obtained in the Hartree-Fock field of one (two) $1s$-vacancies. The junction of the double and wide black lines corresponds to the *overlapping* integral ($\langle xp | yp \rangle$, $\langle 1s | ys \rangle$). Dashed line – Coulomb interaction. A light circle is the top of the radiation transition. The direction of time is from left to right ($t_1 < t_2$). $\hbar\omega$ is the energy of the absorbed photon.

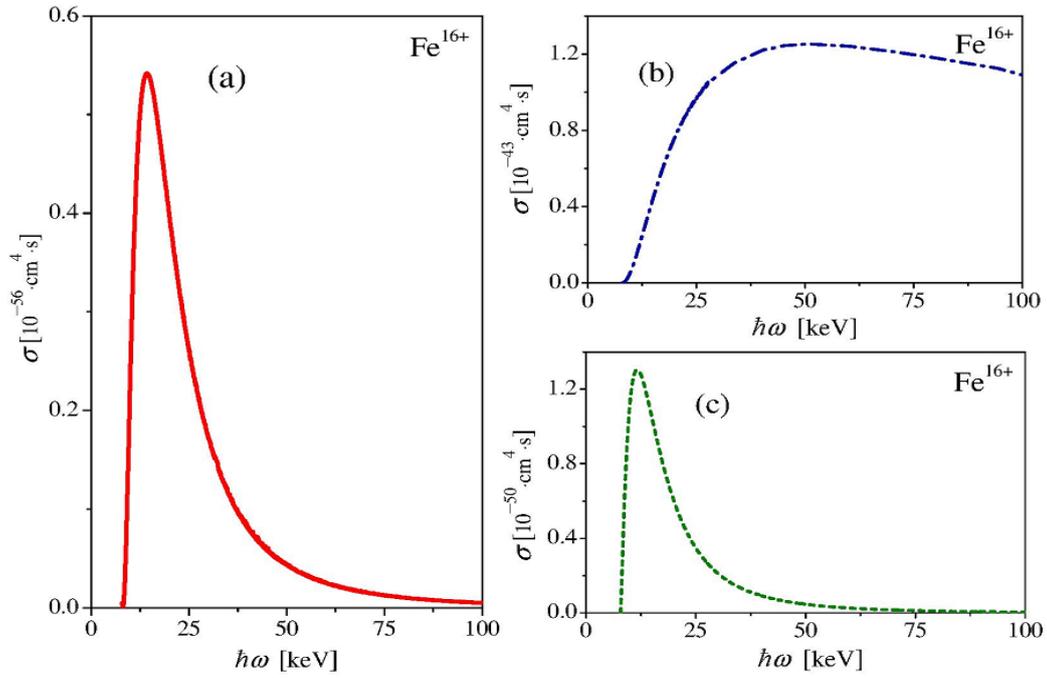

**Figure 2.** Partial generalized cross-sections of the two-photon double ionization of the *K*-shell of the Fe$^{16+}$ ion: a - along the channel (1) from [2] for $l_1 l_2 = pp$ (see Fig. 1a); b (c) – by the channels (1) from [2] for $l_1 l_2 = ss, sd$ (see Fig. 1b) in dipole (non-dipole) approximation for the $\hat{R}$-operator. $\hbar\omega$ is the energy of the absorbed photon.

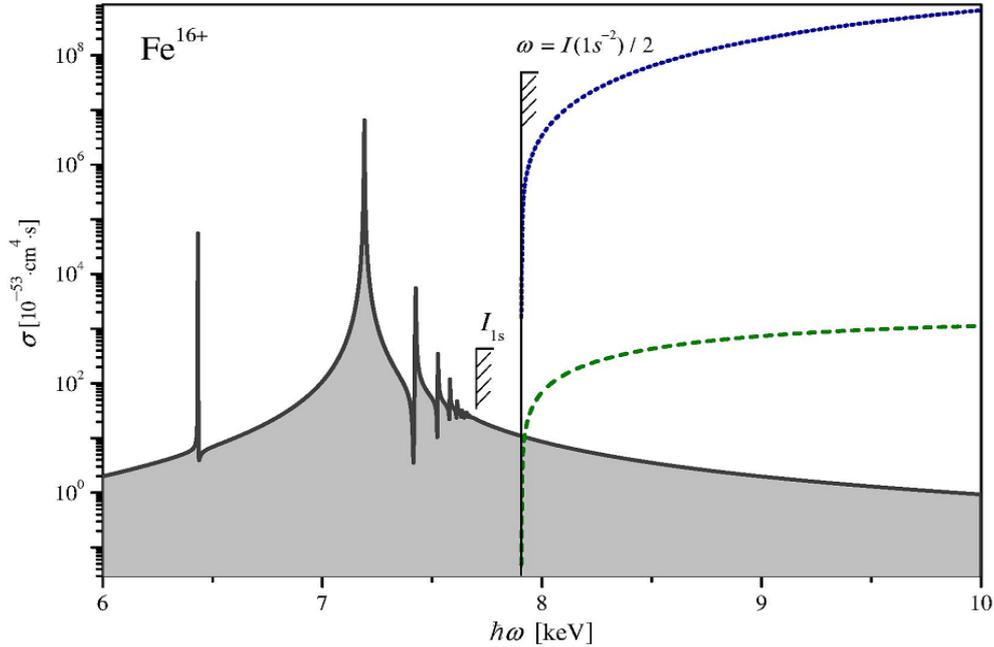

**Figure 3.** The total (by channels (1) from [2]; Fig. 2) generalized cross-section of the two-photon *double* ionization of the *K*-shell of the Fe$^{16+}$ ion: dotted (dashed) curve - dipole (non-dipole) approximation for the $\hat{R}$-operator. A generalized cross-section of a two-photon *single* ionization of the Fe$^{16+}$ ion for $\hbar\omega \in (6; 10)$ keV is given as a *background*. $\hbar\omega$ is the energy of the absorbed photon. $I_{1s}$ is the ionization threshold energy of the 1*s*-shell. $I(1s^{-2})$ is the energy of the threshold for the formation of an *empty K-shell*.